# On shear flow stabilization concepts for the dense z – pinch

F.Winterberg

University of Nevada, Reno

November 2009

**Abstract**

Different ways to achieve the stabilization of a linear z-pinch by a superimposed shear flow are analyzed. They are:

1) Axial shear flow proposed by Arber and Howell with the pinch discharge in its center, and experimentally tested by Shumlak et al.
2) Spiral flow of a dense low temperature plasma surrounding a dense pinch discharge.
3) A thin metallic projectile shot at a high velocity through the center of the pinch discharge.
4) The replacement of the high velocity projectile by the shape charge effect jet in a conical implosion.
5) The replacement of the jet by a stationary wire inside the conical implosion.

**1. Introduction**

The linear z-pinch effect is the oldest and most simple magnetic plasma configuration. But because of its instability it was abandoned a long time ago as a useful configuration for the release of energy by nuclear fusion. Even if stabilized, a pinch discharge as steady state magnetic confinement device would require that the discharge channel must be rather long to keep the end losses low. This requirement is greatly relaxed if it is the goal to ignite a thermonuclear detonation wave propagating at supersonic velocities along the pinch discharge channel. There the time needed to keep the pinch stable is short [1, 2]. The ignition of the detonation wave could be done with a pulsed laser, for example. But for the ignition of the thermonuclear detonation wave, the pinch current must be of the order $10^7$ Ampere to keep the fusion α-particles entrapped in the pinch, and the pinch must have a high density.

The idea to stabilize the pinch discharge by a superimposed velocity shear can be understood from the basics magnetohydrodynamic equations for a plasma of infinite conductivity and zero viscosity:

$$\rho \frac{\partial \mathbf{v}}{\partial t} = -\text{grad}\, p - \frac{\rho}{2} \text{grad}\, \mathbf{v}^2 + \rho \mathbf{v} \times \text{curl}\mathbf{v} - \frac{1}{4\pi} \mathbf{B} \times \text{curl}\mathbf{B} \qquad (1)$$

to be supplemented by:

$$\frac{\partial \mathbf{B}}{\partial t} = \text{curl}(\mathbf{v} \times \mathbf{B}) \qquad (2)$$

For a steady state solution ∂/∂t = 0, hence curl**v**×**B** = 0, possible only if **v**∥**B** or if there is a large electric field **E** = (-1/c)**v**×**B**.

In the absence of a vortex flow where curl**v** = 0, the usual magnetohydrodynamic instabilities arise from the last term on the r.h.s of eq. (1). This means, these instabilities should be suppressed if the flow has vorticity and that

$$\rho \mathbf{v} \times \text{curl}\mathbf{v} > \frac{1}{4\pi} \mathbf{B} \times \text{curl}\mathbf{B} \tag{3}$$

For $\rho$ = const, this becomes

$$\mathbf{v} \times \text{curl } \mathbf{v} > \mathbf{v}_A \times \text{curl } \mathbf{v}_A \tag{4}$$

where, $\mathbf{v}_A = \mathbf{B}/\sqrt{4\pi\rho}$ is the Alfven velocity.

Introducing the Alfven Mach number $M_A = v/v_A$, the inequality (4) then simply means that

$$M_A > 1 \tag{5}$$

Because the stabilizing term $\rho \mathbf{v} \times \text{curl } \mathbf{v}$ depends both on $\rho \mathbf{v}$ and curl **v**, the wording shear flow stabilization is somewhat misleading. A better way to describe it would be to call it dense vorticity flow stabilization.

Now, if somehow a high temperature but lower density magnetized pinch discharge plasma can be separated from a low temperature but high density non-magnetized vortex flow, for example by placing the pinch discharge inside the hollow core of a high density line vortex of infinite conductivity, the pinch discharge would be "wall stabilized" with regard to the wall of vortex core. In this case the condition (5) is replaced by

$$\rho_0 v_0^2 > nkT \tag{6}$$

where $\rho_0$ and $v_0$ are the density and velocity of the vortex flow. If for the pinch $\beta = (nkT)/(B^2/8\pi) \simeq 1$, one has $v_A^2 = kT/M_H$, where $M_H$ is the proton mass, thus $M_A = v_0/v_A$. Then setting $\rho_0 = n_0 A M_H$, where $n_0$ and A are the particle number density and atomic weight in the vortex flow, one obtains

$$A\left(\frac{n_0}{n}\right) M_A^2 > 1 \tag{7}$$

or that

$$M_A > \sqrt{\frac{n}{An_0}} \tag{8}$$

For $M_A > 1$ the stabilizing flow can here be even subsonic. If for example $n_0/n \simeq 10$, and $A = 40$ (argon plasma), then $M_A > 1/\sqrt{400} = 1/20$. Since for a fusion plasma $v_A \simeq 10^8$ cm/sec one finds that for $M_A > 1/20$, one has $v_0 > 5 \times 10^6$ cm/s. Or still better, for $A \simeq 200$ (lead plasma) and $n_0/n \simeq 10$, one has $M_A > 1/45$, or $v_0 > 2 \times 10^6$ cm/s. This demonstrates the usefulness of separating the hot fusion plasma from a cool stabilizing flow. This however, is only possible if there is a heat insulating surface, a vacuum or a low density plasma layer, in between the hot fusion plasma and the surrounding much cooler but stabilizing flow.

## 2. The Arber-Howell and Shumlak et al. configurations

A configuration proposed by Arber and Howell, assumes that an axial plasma flow can be superimposed onto the pinch [3]. This idea was further analyzed, both theoretically and experimentally in a series of papers by Shumlak et al. [4, 5, 6]. It is there argued, that the magnetohydrodynamic equilibrium equation (1), setting $\partial/\partial t = 0$, is unchanged by an axial flow. This however, is not quite correct, since it does not leave unchanged eq. (2), where, unless **v** || **B**, a radial electric field

$$\mathbf{E} = -\frac{1}{c}\mathbf{v} \times \mathbf{B} \ [\text{esu}] \tag{9}$$

is set up by the axial flow, ignored in the stability analysis.

For the dense z-pinch, with a current of $10^7$ Ampere and a millimeter size diameter, on has B $\simeq 10^7$ Gauss. To get for a DT z-pinch fusion plasma $M_A>1$, v must be larger than the thermal velocity $v_{th} \simeq 10^8$ cm/s. To get, $M_A>1$ one finds that E $>10^{-2}$ B = $10^5$esu$\simeq 10^7$ Volt/cm. Under these circumstances a small helical distortion of the z-pinch channel will lead to an axial magnetic field $B_z$, which together with a radial current will lead to an azimuthal force on the pinch discharge, resulting in its rotation.

In the experiment done by Shumlak et al., a large axial flow was achieved by attaching a z-pinch channel to the end of co-axial plasma gun. Because of the large axial flow, no large radial plasma implosion, typical for the plasma focus, did there occur. But in one experiment a hollow pinch discharge channel was observed, speaking for at least some rotation [5]. In all of these experiments the pinch current was far away from the $10^7$ Ampere required for detonation, and the density of the pinch comparatively small.

## 3. Improving the Shumlak-Configuration

Taking into account our foregoing analysis, we suggest the following improvement of the Shumlak (ZaP) pinch configuration: 1. The z-pinch discharge is separated from the axial shear flow by making the z-pinch discharge from a separate breakdown between the two electrodes of the linear z-pinch discharge. 2. The axial shear flow surrounding the stationary z-pinch plasma is an argon plasma or some other high A- number plasma, accelerated in the co-axial plasma gun. In this configuration, the high velocity-low temperature, but dense high A-number plasma is separated from the stationary low density but high temperature z-pinch plasma, with both of them separated by a boundary layer. To prevent mixing of both components, an axial magnetic field is applied to the barrel of the plasma gun, resulting in a helical motion the high A-number plasma, where the centrifugal force tends to separates the high A-number plasma from the pinch plasma. Furthermore, the thermomagnetic currents set up by the Nernst effect in the boundary layer between both, will repel them from each other (see Appendix).

Because for a high density $\rho$, the $\rho \mathbf{v} \times \text{curl} \mathbf{v}$ term can be much larger than the $(1/4\pi)\mathbf{B} \times \text{curl}\mathbf{B}$ term, and it is here not necessary to have for stabilization an axial flow with a velocity in excess of $\sim 10^8$ cm/s.

## 4. The needle-like projectile shot through the pinch core configuration

This configuration shown in **Fig.4** is actually the first shear flow configuration proposed [7, 8]. There the density of the flowing metallic material can be of the order 10gm/cm$^3$, with the pinch at rest. If moving with a velocity of $3 \times 10^5$cm/s (attainable with a H$_2$-O$_2$ gun), one has a stagnation pressure equal to $\rho v^2 \sim 10^{12}$ dyn/cm$^2$, which can for the pinch discharge withstand a magnetic field of $\sim 10^7$ Gauss. Instead of a H$_2$-O$_2$ gun, one may use a light gas gun to bring the needle to a velocity of $\sim 10^6$cm/s. There the pressure is $10^{13}$dyn/cm$^2$, that is ten times higher, whereby $B^2/8\pi = 10^{13}$ dyn/cm$^2$, with B well above $10^7$ Gauss. With a magnetic travelling wave accelerator finally, the needle can be accelerated to 30km/s, with $B^2/8\pi \simeq 10^{14}$ dyn/cm$^2$, hence B$\simeq$5$\times$10$^7$ Gauss (**Fig. 5**).

In the thin needle through the pinch core approach, a large $\rho\mathbf{v} \times$ curl $\mathbf{v}$ is possible without stretching technology. The pinch is there hollow and surrounds the projectile, but because of the small radius of the needle, the pinch can have a large density, with the Nernst effect magnetically insulating the cold needle against the hot pinch plasma [see Appendix].

**5. Conical Wire Array Implosion Shape Charge Effect Configuration**

Instead of accelerating and shooting a thin needle through a pinch discharge channel, one may instead make use of the well known shape charge effect as shown **Fig .6**. There the conical implosion is achieved by the electric pulse power implosion of a conical wire array. This configuration is easier to realize, but less suitable for a "rep-rated" operation, because the wire array is destroyed after each shot.

**6. Conical Wire Array Implosion with Central Wire**

Replacing the jet with a central wire one obtains another configuration (**Fig.7**). The idea here is to stabilize the central wire against the pinch instabilities by an axial plasma flow generated by the conical implosion, producing shear along the central wire. It is somehow the complementary configuration to the needle shot through the pinch, where the wire (needle) is moving and the plasma is at rest.

For the stationary central wire configuration one must have

$$(\rho v^2)_{plasma} > B^2/4\pi \qquad (10)$$

or

$$M_A > 1 \qquad (11)$$

This means if for a fusion plasma where $v_A \simeq 10^8$ cm/s, then $v > 10^8$ cm/s. This large velocity is difficult to realize.

## **7. Discussion**

At low pinch densities the modified Shumlak et al. configuration, separating the dense pinch plasma from the spiraling high density shear flow, has a distinct advantage because it does not require a supersonic shear flow. It also permits to reach a higher z-pinch density, because there no electric field is set up by the **v**×**B** term. But for a non-pulsed steady state configuration it still would require a rather large pinch column to reduce the end losses. Against the modified Shumlak et al. configuration, stands the unavoidable mixing of the high A-number spiraling shear flow with the non-flowing z-pinch plasma. It is for this reason, that the steady state configuration proposed by Hassam and Huang [9] has to be preferred.

Comparing the stationary central wire surrounded by fast axially moving pinch discharge, configuration, with the complementary fast moving central wire through a stationary pinch configuration, the fast moving central wire configuration is the clear winner, because it only requires a subsonic velocity of the wire in the >10km/s range, instead of the more than $10^3$km/s dense plasma flow required for the stationary wire configuration.

This result is surprising only if one incorrectly thinks that it should make no difference if the wire is at rest with the plasma flowing, or if the plasma is at rest with the wire moving. In reality the wire at rest is attached to the large mass of the earth, and vice verse the pinch plasma at rest is part of an apparatus attached to earth. In each case this is a two body problem, where one of the masses is very large. Therefore, in the first case it is the momentum flux density $\rho v^2$ of the moving plasma against the earth, and in the second case the momentum flux density $\rho_0 v_0^2$ of the moving wire against the earth.

For a dense z-pinch at thermonuclear temperatures the density is $\rho \sim 10^{-3}$ g/cm$^3$, and the velocity of sound a~$10^8$ cm/s. For $M_A \simeq 1$, one there has $\rho v^2 \sim 10^{13}$ erg/cm$^3$. This can be reached with a metallic wire of density $\rho_0 \sim 10$ g/cm$^3$ moving with 10km/s=$10^6$ cm/s. But it much easier to accelerate a wire to 10km/s, by a travelling magnetic wave accelerator, for example, than a dense plasma to more than 1000 km/s, requiring to heat an expanding plasma to more than $10^8$ °K [11].

**Appendix**

If a magnetic field exists in between two media one hot the other one cold, with a gradient in between, electric currents are induced in the boundary layer between both by the thermomagnetic Nernst effect, acting here like a self exciting thermomagnetic dynamo requiring not more than a initial magnetic seed field.

The current density by the thermomagnetic Nernst effect is [9]

$$\mathbf{j}_N = \frac{3nkc}{2H^2} \mathbf{B} \times \nabla T, \qquad (A.1)$$

With the magnetic force density acting on the jet plasma in the boundary layer,

$$\mathbf{f} = \frac{1}{c}\mathbf{j}_N \times \mathbf{B} = \frac{3nkc}{2H^2}(\mathbf{B} \times \nabla T) \times \mathbf{B} \qquad (A.2)$$

Since $\nabla T \perp \mathbf{B}$ this becomes

$$\mathbf{f} = \frac{3}{2}nk\nabla T \qquad (A.3)$$

The magnetohydrostatic equation in the boundary layer is

$$\nabla p = \mathbf{f} \qquad (A.4)$$

and with $p = 2nkT$,

$$\nabla p = 2nk\nabla T + 2kT\nabla n = \frac{3}{2}nk\nabla T \qquad (A.5)$$

it becomes

$$n\nabla T + 4T\nabla n = 0 \qquad (A.6)$$

which upon integration yields

$$Tn^4 = T_0 n_0^4 \tag{A.7}$$

where $T_0$ and $n_0$ are temperature and particle number in the boundary layer. This means, unlike for constant pressure where nT = const, the Nernst effect with $n^4$T = const, repels the the cold flow from the hot pinch.

**References**

1. F. Winterberg, in "Physics of High Energy Density", Academic Press, New York 1971, p.376.
2. F.Winterberg, Atomkernenergie **39**, 265 (1981).
3. T. D. Arber and D. H. Howell, Phys. Plasma **3**, 554 (1969).
4. U. Shumlak and C. W. Hartman, Phys. Rev. Lett. **75**, 3285 (1995).
5. U. Shumlak et al., 2002 IAEA, FEC paper, EX/P1-19.
6. U. Shumlak et al., Journal of Fusion Energy, **26**,185 (2006).
7. F. Winterberg, Atomkerneregie-Kerntechnik **43**, 31 (1983).
8. F. Winterberg, Beitr. Plasmaphysics **25**, 117 (1985).
9. A. B. Hassam and Yi-Min Huang, Phys. Rev. Lett. **91**, 195002-1 (2003).
10. L. Spitzer, Physics of Fully Ionized Gases, Interscience Publishes, John Wiley & Sons, New York 1962, p. 145.
11. Proceedings of the Impact Fusion Workshop, Los Alamos, NM., July 10-12, 1979, Editor A. T. Peaslee Jr., LA-8000-C Conference UC-21, August 1979.

**Figures**

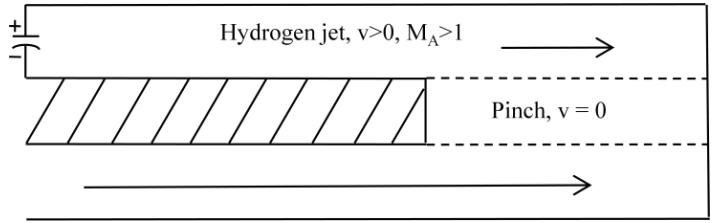

**Fig.1.** Original ZaP configuration.

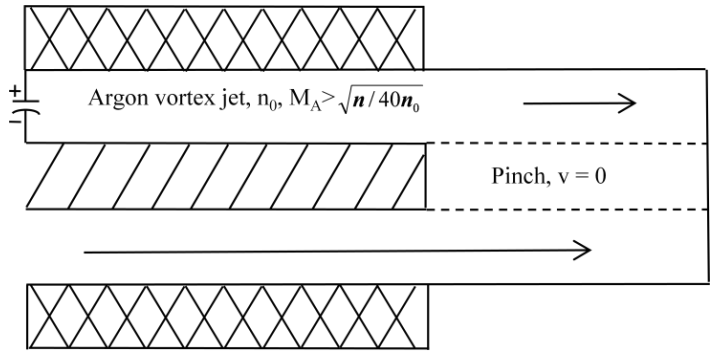

**Fig.2.** 1st Modified ZaP configuration.

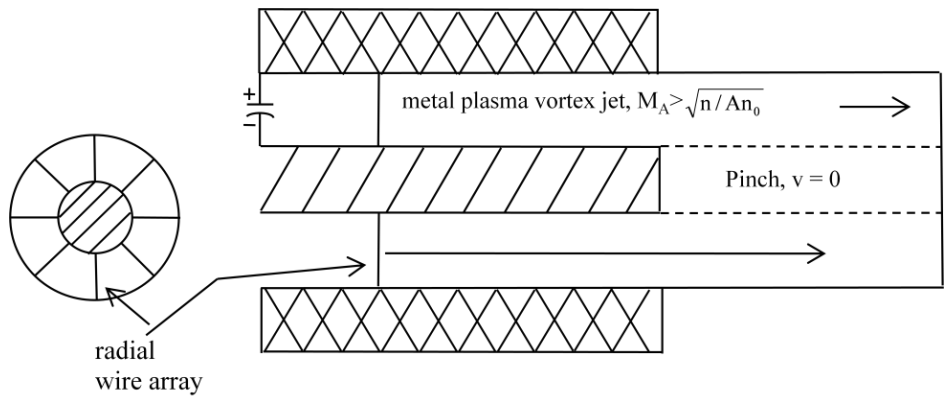

**Fig. 3.** 2nd Modified ZaP configuration.

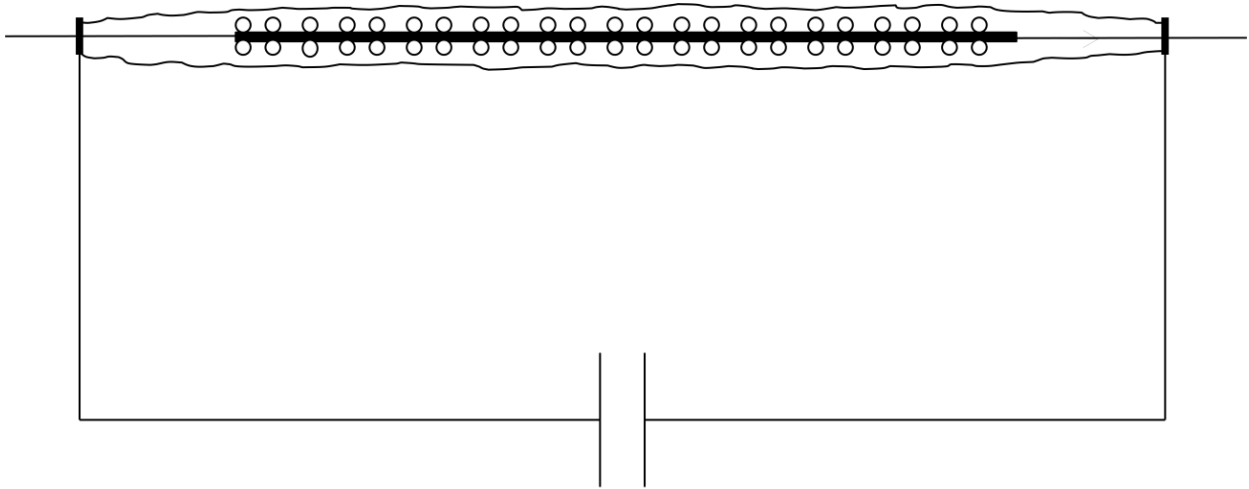

**Fig. 4.** Shear flow vorticity stabilization with high velocity injection of a metallic jet through the core of a pinch discharge.

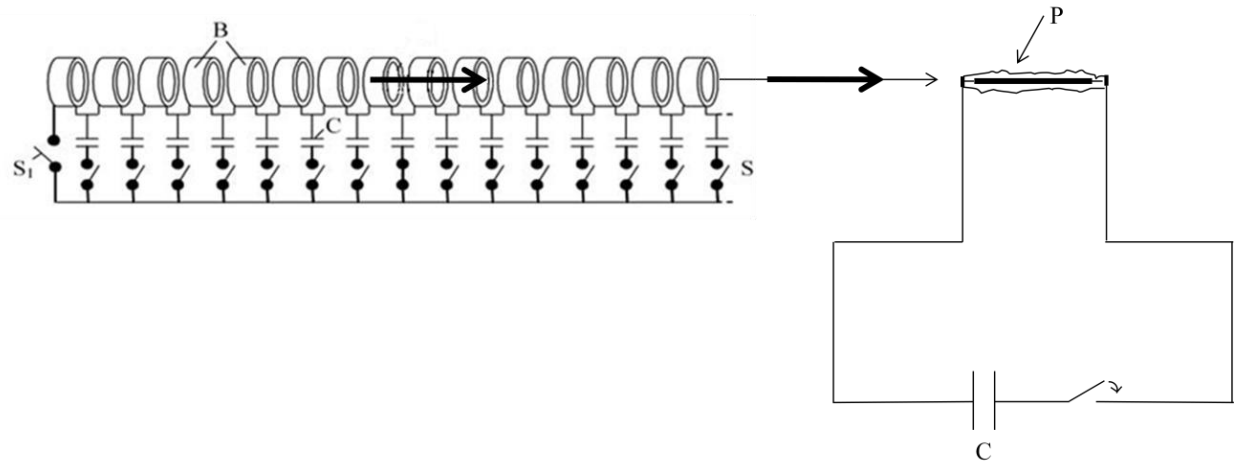

**Fig.5.** Hypervelocity acceleration and injection of magnetized plasma cylinder onto a linear pinch discharge. B magnetic field coil; C large capacitor banks; S and S1 switches; P pinch discharge.

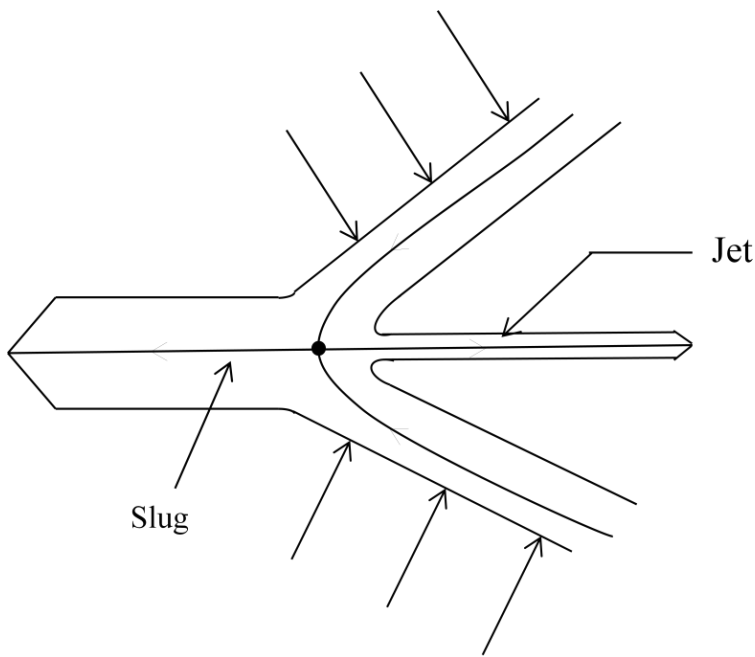

**Fig.6.** Jet formation in a conical implosion.

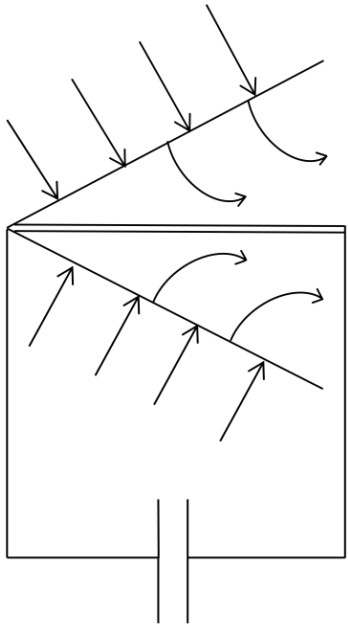

**Fig.7.** Conical implosion of plasma onto a central rod over which a pinch discharge takes place.